\documentclass[%
    reprint,
    showpacs,preprintnumbers,
    amsmath,amssymb,
    aps,
    prl,
]{revtex4-1}

\usepackage{amsmath}

\def\be#1{\begin{equation}\label{#1}}
\def\ee{\end{equation}}
\newcommand {\ba}[2]{\be{#1}\begin{array}{#2}}
\newcommand {\ea}{\end{array} \ee}
\def\eq#1{(\ref{#1})}
\newcommand{\qq}{\,,\qquad}
\renewcommand{\=}{\stackrel{\mbox{\scriptsize def}}{=}}
\let\TS=\textstyle
\let\DS=\displaystyle
\def\({\left(}
\def\){\right)}
\def\aav#1{\langle{#1}\rangle}
\let\w = \omega
\def\kB{k_{\!B}}
\def\k{k}
\def\pt#1{}
\def\fig#1{Fig.~#1}

\usepackage{ifpdf}
\ifpdf
  \usepackage[pdftex]{graphicx}
  \DeclareGraphicsExtensions{.png,.jpg,.tif}
\else
  \usepackage{graphicx}
  \DeclareGraphicsExtensions{.bmp,.pcx}
\fi

\newlength{\pix}
\begin{document}

\title{On Unsteady Heat Conduction in a Harmonic Crystal}

\author{Anton M. Krivtsov}\email{akrivtsov@bk.ru}
\affiliation{%
 Peter the Great Saint Petersburg Polytechnic University, \\
 Institute for Problems in Mechanical Engineering RAS
}%

\date{November 22, 2015}

\begin{abstract}
An analytical model of unsteady heat transfer in a one-dimensional harmonic crystal is presented. A nonlocal temperature is introduced as a generalization of the kinetic temperature. A closed equation determining unsteady thermal processes in terms of the nonlocal temperature is derived. For an instantaneous heat perturbation a time-reversible equation for the kinetic temperature is derived and solved. The resulting constitutive law for the heat flux in the considered system is obtained. This law significantly differs from Fourier's law and it predicts a finite velocity of the heat front and independence of the heat flux on the crystal length. The analytical results are confirmed by computer simulations.
\end{abstract}


\maketitle

An understanding of heat transfer at microlevel is essential to obtain link between microscopic and macroscopic description of solids~\cite{Hoover 2015, Charlotte 2012}.
As far as macroscopic scale level is concerned the Fourier law of heat conduction is widely and successfully used to describe heat transfer processes. At microscopic level, however, analytical and numerical investigations have shown substantial deviations from Fourier's law~\cite{Bonetto 2000, Lepri 2003, Dhar 2008}. These inadequacies can be on principle addressed by using
special laws of particles interactions~\cite{Aoki 2000, Gendelman 2000, Giardina 2000, Gendelman 2014} or complex enough structures~\cite{Bonetto 2004, Le 08 DAN}. Recent experimental data however showed that Fourier's law is indeed violated in low-dimensional nanostructures~\cite{Chang-Zettl 2008, Xu 2014, Hsiao 2015}. This motivates interest to the simplest lattice models, in particular harmonic one-dimensional crystals (chains), where these anomalies are most prominent~\cite{Kannan 2012, Dhar 2015}.
Problems of this kind previously have been mainly addressed in the context of the steady-state heat conduction~\cite{Bonetto 2000, Lepri 2003, Dhar 2008}.
The present work focuses on unsteady conduction regimes~\cite{Le 08 DAN, Rubin 1963, Gendelman 2012, Gusev-Lurie 2012}.

Here we suggest an approach that allows rigorous derivation of macroscopic heat conduction equations and corresponding anomalous heat conduction law for harmonic systems in a one-dimensional, non-quantum case. The obtained equations differ substantially from the earlier suggested heat transfer equations~\cite{Chandrasekharaiah 1986, Tzou 2015}, however they are in excellent agreement with molecular dynamics simulations and previous analytical estimations~\cite{Gendelman 2012}.
The approach presented here is developed in context of one-dimensional systems, however the same ideas can be applied to systems of higher dimensions.

Thus we consider a one-dimensional crystal, described by the following equation of motion:
\be{1}
     \ddot{u}_i = \w_0^2(u_{i-1}-2u_i+u_{i+1})
     \qq \w_0 = \sqrt{C/m},
\ee
where
 $u_i$ is the displacement of the $i$th particle,
 $m$ is the particle mass,
 $C$ is the stiffness of the interparticle bond.
The crystal is infinite: the index $i$ is an arbitrary integer.
The initial conditions are
\be{2}
     u_i|_{t=0} = 0
     \qq
     \dot u_i|_{t=0} = \sigma(x)\varrho_i,
\ee
where $\varrho_i$ are independent random values with zero expectation and unit variance; $\sigma^2(x)$ is variance of the initial velocities, which is a slowly varying function of the spatial coordinate $x=ia$, where~$a$ is the lattice constant. These initial conditions correspond to an instantaneous temperature perturbation, which can be induced in crystals, for example, by an ultrashort laser pulse~\cite{Poletkin 2012}. The displacements as functions of time $u_i = u_i(t)$ can be found as a solution of the Cauchy problem \eq{1}--\eq{2}. In addition, these functions are linearly dependent on the integration constants, which are random due to the random nature of the initial conditions~\eq{2}.

The first analytical solution of a steady heat conduction problem for a harmonic chain was obtained in~\cite{Rieder 1967} using a covariance matrix for coordinates and momenta. Then this approach was extended and applied to various harmonic systems~\cite{Bonetto 2000, Bonetto 2004, Kannan 2012, Dhar 2015}. Study of the covariance matrix allowed obtaining analytical expressions for steady~\cite{Lepri 2009} and unsteady~\cite{Lepri 2010, Delfini 2010} temperature profiles.
In this paper a somewhat similar approach based on analysis of covariances for velocities~\cite{Krivtsov 06 APM, Krivtsov 2014 DAN, Krivtsov 2015 DAN} is used. We introduce a {\it nonlocal temperature}~$\theta_n(x)$ as
\be{4}
    \kB(-1)^n \,\theta_n(x) \= m\aav{\dot u_i \dot u_j}
,\ee
where $\kB$ is the Boltzmann constant,
$n=j-i$ is the covariance index, $x=\frac{i+j}2a$ is the spatial coordinate, angle brackets stand for mathematical expectation, $\aav{\dot u_i \dot u_j}$~is the velocity covariance (note that $\aav{\dot u_i}\equiv\aav{\dot u_j}\equiv0$). If ${n=0}$ then $i=j$ and quantity $\theta_n$ coincides with the kinetic temperature $T$: $\theta_0(x) = T(x) = \frac m{\kB}\aav{\dot u_i^2}$, where $i=x/a$. According to its definition, the introduced nonlocal temperature reflects a nonlocal nature of thermal processes in harmonic crystals and can be considered as a generalization of the kinetic temperature.

Let us calculate the second time derivative of $\theta_n$ using the dynamics equations \eq{1} and the following two approximations.
\begin{enumerate}

\item
The nonlocal temperature $\theta_n(x)$ is a slowly varying function of the spatial coordinate~$x$ (on the distances of order of the lattice constant~$a$). This allows replacing the finite differences by spatial derivatives~\cite{Born 1958}. The approximation is adequate for processes that are sufficiently smooth in space, e.~g. for spatial temperature profiles in a form of waves that are much longer then $a$.

\item
The virial approximation~\cite{Hoover 2015}: time or spatial derivatives of mathematical expectations are small with respect to quantities that have non-zero values in thermodynamic equilibrium. In particular, this allows us to express covariances of the bond strains in terms of the nonlocal temperature. The approximation is adequate for processes that are not too far from thermodynamic equilibrium.

\end{enumerate}
Then after some transformations we obtain a differential-difference equation
\be{7}\TS
    \ddot\theta_n + \frac14c^2(\theta_{n-1}-2\theta_{n}+\theta_{n+1})'' = 0,
\ee
where $c=\w_0 a$ is the speed of sound.
This is a closed equation describing unsteady thermal processes in the crystal in terms of the nonlocal temperature. The processes under consideration should be such that the nonlocal temperature is sufficiently smooth in time and space. Apart from this limitation any unsteady thermal processes satisfy equation \eq{7}. After solution (analytical or numerical) of equation \eq{7} the kinetic temperature can be obtained as $T(t,x) = \theta_n(t,x)|_{n=0}$.

The initial conditions for equation~\eq{7} corresponding to the original initial conditions~\eq{2} are:
\be{8}
    \theta_n|_{t=0} = T_0(x)\delta_n \qq
    \dot\theta_n|_{t=0} = 0 ,
\ee
where
$T_0(x)=\frac1{2\kB}m\sigma^2(x)$
is the initial temperature distribution;
${\delta_{n} = 1}$ for $n=0$ and ${\delta_{n} = 0}$ for $n\ne 0$.
The initial conditions \eq{8} are taken after a fast transition process, which results, according to the virial theorem, in a double reduction of the initial kinetic temperature~\cite{Krivtsov 2014 DAN}.
Note that in contrast with the random initial value problem \makebox{\eq{1}--\eq{2}}, the initial value problem \makebox{\eq{7}--\eq{8}} is expressed in terms of mathematical expectations, and therefore it is a deterministic problem.

Using an integral Fourier transform in the spatial coordinate~$x$ the problem \eq{7}--\eq{8} can be solved analytically. For the Fourier image~$\hat\theta_n(t,\k)$ we obtain
\ba{9}{c}
    \ddot{\hat\theta}_n = \frac14c^2\k^2(\hat\theta_{n-1}-2\hat\theta_{n}+\hat\theta_{n+1}) , \\[2mm]
    \hat\theta_n|_{t=0} = \hat T_0(\k)\delta_n \qq
    \dot{\hat\theta}_n|_{t=0} = 0,
\ea
where $\k$ is the spatial frequency, $\hat T_0(\k)$ is the Fourier image of the initial temperature distribution~$T_0(x)$. Let us note the similarity between \eq{1}--\eq{2} and \eq{9}: initial value problem \eq{9} can be interpreted as a motion of a harmonic chain having an initial shift of the central particle. This kind of problems can be effectively solved in terms of Bessel functions. In particular, Bessel functions were successfully applied to solution of shock-wave problems in harmonic chains~\cite{Manvi 1969, Holian 1978}. Similarly, the problem~\eq{9} has an analytical solution
$\hat\theta_n(t,\k) = \hat T_0(\k)J_{2n}(c\k t)$, where $J_{2n}$ are the Bessel functions of the 1st kind~\cite{Abramovits 1979}.
From the practical point of view the most interesting case is $n=0$, which gives Fourier image $\hat T(t,\k)$ of the kinetic temperature distribution:
\be{11}
    \hat T(t,\k) = \hat T_0(\k)J_{0}(c\k t).
\ee
From \eq{11} it follows that the image $\hat T(t, \k)$
satisfies the Bessel differential equation
\be{12}
    \ddot{\hat T} + \frac1t\dot{\hat T} = -c^2\k^2\hat T.
\ee
Fourier inversion of \eq{12} gives a partial differential equation for the temperature field
\be{13}
    \ddot T + \frac1t\dot T = c^2 T''.
\ee
The corresponding initial conditions follow from \eq{8}:
\be{14}
    T|_{t=0} = T_0(x) \qq
    \dot T|_{t=0} = 0.
\ee
Fourier inversion of the representation \eq{11} gives an analytical solution of the initial value problem \eq{13}--\eq{14}:
\be{15}
    T(t,x) = \frac1\pi\int_{-t}^t\frac{T_0(x-c\tau)}{\sqrt{t^2-\tau^2}}\,d\tau.
\ee
Similar integral representation was obtained in \cite{Shoby 1981} using heat energy density correlation functions.

Thus, the evolution of the temperature field in a one-dimensional crystal after an instantaneous thermal perturbation is described by partial differential equation \eq{13} with initial conditions~\eq{14} or by integral formula~\eq{15}. According to \eq{15} the thermal front propagates with the sound speed~$c$ (in contrast to the thermal conductivity based on Fourier's law where an unphysical instantaneous signal propagation is realized).
The obtained wave behavior of the heat front is similar to predictions of the wave theories of heat conduction~~\cite{Chandrasekharaiah 1986, Tzou 2015}. However, the obtained solution has important differences, which will be shown in the text to follow.

Let us consider the heat flux. For the considered system it can be represented~\cite{Lepri 2003, Dhar 2008, Hoover 1986} as
\be{15a}\TS
    q = \frac12C\aav{(u_{i} - u_{i+1})(\dot u_{i} + \dot u_{i+1})}.
\ee
The heat flux $q$ satisfies the energy balance equation
\be{16a}
    \rho \kB\dot T = - q',
\ee
where $\rho=1/a$ is the density (number of particles per unit volume).
Joint consideration of equations~\eq{13} and~\eq{16a} gives the constitutive law for the heat flux
\be{16}
    \dot q + \frac1t q = - \kB\rho c^2 T',
\ee
which replaces Fourier's law in the considered system.
Alternatively, the law \eq{16} can be derived directly, in the same way as equation~\eq{13} is derived. An integral representation for the heat flux follows from~\eq{15} and~\eq{16}:
\be{17}
    q(t,x) = \frac{\kB\rho c}{\pi t}\int_{-t}^t\frac{T_0(x-c\tau)}{\sqrt{t^2-\tau^2}}\,\tau d\tau.
\ee

Comparison of the obtained heat transfer equation \eq{13} with the heat equation based on Fourier's law and the thermal wave equation based on the Maxwell-Cattaneo-Vernotte (MCV) law is given in table~1.
\begin{table}[htb]
\def\ll{30mm}
\begin{tabular}{c|c|c|c}
&
\parbox{18mm}{\centering Heat equation \\ (Fourier)} &
\parbox{27mm}{\centering Thermal wave equation \\ (MCV)} &
\parbox{32mm}{\centering Equation \eq{13} \\ (present paper)}
\\ [5mm] \hline
&&\\ [-1mm]
a) &
$\dot T = \beta T''$ &
$\DS\ddot T + \frac1\tau\dot T = \frac\beta\tau T''$ &
$\DS\ddot T + \frac1t\dot T = c^2 T''$
\\ [8mm]
b) &
$q = -\kappa T'$ &
$\DS\dot q + \frac1\tau q = -\frac\kappa\tau T'$ &
$\DS\dot q + \frac1t q = - \kB\rho c^2 T'$
\\ [8mm]
c) &
$e^{-\beta\k^2 t}$ &
$\approx\,e^{-\frac t{2\tau}}\cos{(\k c t)}$ &
$J_0(\k ct) \approx \frac{\cos{\(\k ct-\frac\pi4\)}}{\sqrt{\frac\pi2\k c\,t}}$
\\ [-3mm] &&\\
\end{tabular}
\caption{%
a) Heat transfer equation,
b) equation connecting heat flux and temperature,
c) decay law for the sinusoidal heat perturbation.
Notations:
$t$ is time (variable),
$\tau$ is the relaxation time (constant),
$\beta$ is the thermal diffusivity,
$\kappa$ is the thermal conductivity,
$c$ is the sound speed,
$\rho$ is the density,
$\kB$ is the Boltzmann constant,
$k$ is the spatial frequency.
Approximation~(c) for the thermal wave equation is obtained for $c^2=\beta/\tau$ and large $\k$; approximation for $J_0$ is valid for relatively large~$t$.}
\end{table}
The latter equation and equation \eq{13} have similar form and somewhat similar behavior (e.~g. a finite velocity of the heat front propagation). However, there are significant differences. The first one is that $\tau$, a material constant, is replaced in~\eq{13} by the physical time $t$. The second difference is time-reversibility of equation \eq{13}: the equation is not changing when $t$ is replaced by $-t$, same as the original microscopic equation \eq{1}. On contrary, both classical and wave equations of heat conduction are irreversible. The contradiction between time-reversibility of the classical microscopic equations and irreversibility of the corresponding macroscopic continuum equations is one of the opened questions of the modern physics~\cite{Holian 1987, Hoover 2015}. The obtained reversible macroscopic equation of heat conduction may be a step towards solution of this problem.

We consider now a sinusoidal temperature perturbation~\cite{Le 08 DAN, Gendelman 2010}:
\be{o3}
    T_0(x) = A_0\sin{\k x} + B,
\ee
where $A_0$ and $B$ are temperature constants, ${\k = 2\pi/\lambda}$ is the spatial frequency, $\lambda$ is the wavelength of the perturbation.
Formulas \eq{15} and~\eq{17} give exact analytical solution for the temperature and heat flux
\ba{s1}{l}
    T(t,x) = A_0J_0(\k ct)\sin{\k x} + B, \\[3mm]
    q(t,x) = -\kB\rho c A_0J_1(\k ct)\cos{\k x},
\ea
where $J_0$ and $J_1$ are the Bessel functions of the 1st kind.
Previously, an existence of a Bessel function solution for this problem was mentioned in~\cite{Gendelman 2012}, and solution similar to \eq{s1} for the temperature field was obtained in Master-degree thesis~\cite{Kachman 2011}.

To justify the assumptions in derivation of the analytical solution we compare it with results of molecular dynamics (MD) simulations. Equations~\eq{1} are solved by the central differences method, the time step is $0.01 \tau_0$, where ${\tau_0=2\pi/\w_0}$. The initial conditions \eq{o3} are set by a random number generator, the wavelength $\lambda$ is equal to the length of the chain containing $10^4$ particles. To provide correspondence with the analytical approach used above, $10^4$ realizations of such chain with an independent random initiation are computed. To optimize the computations all chains are joined at end-points to form a long chain ($10^8$ particles) with periodic boundary conditions. The results of the computations are compared with analytical solution~\eq{s1} in \fig1.
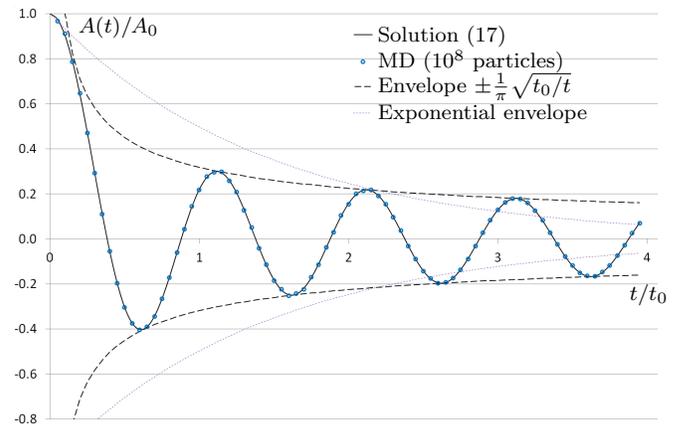
\begin{figure}[htb]\footnotesize
\unitlength = 1.35\pix
\noindent
\begin{picture}(1493,981)(0,0)
\put(0,0){\includegraphics[width = 85.46mm, height = 56.15mm]{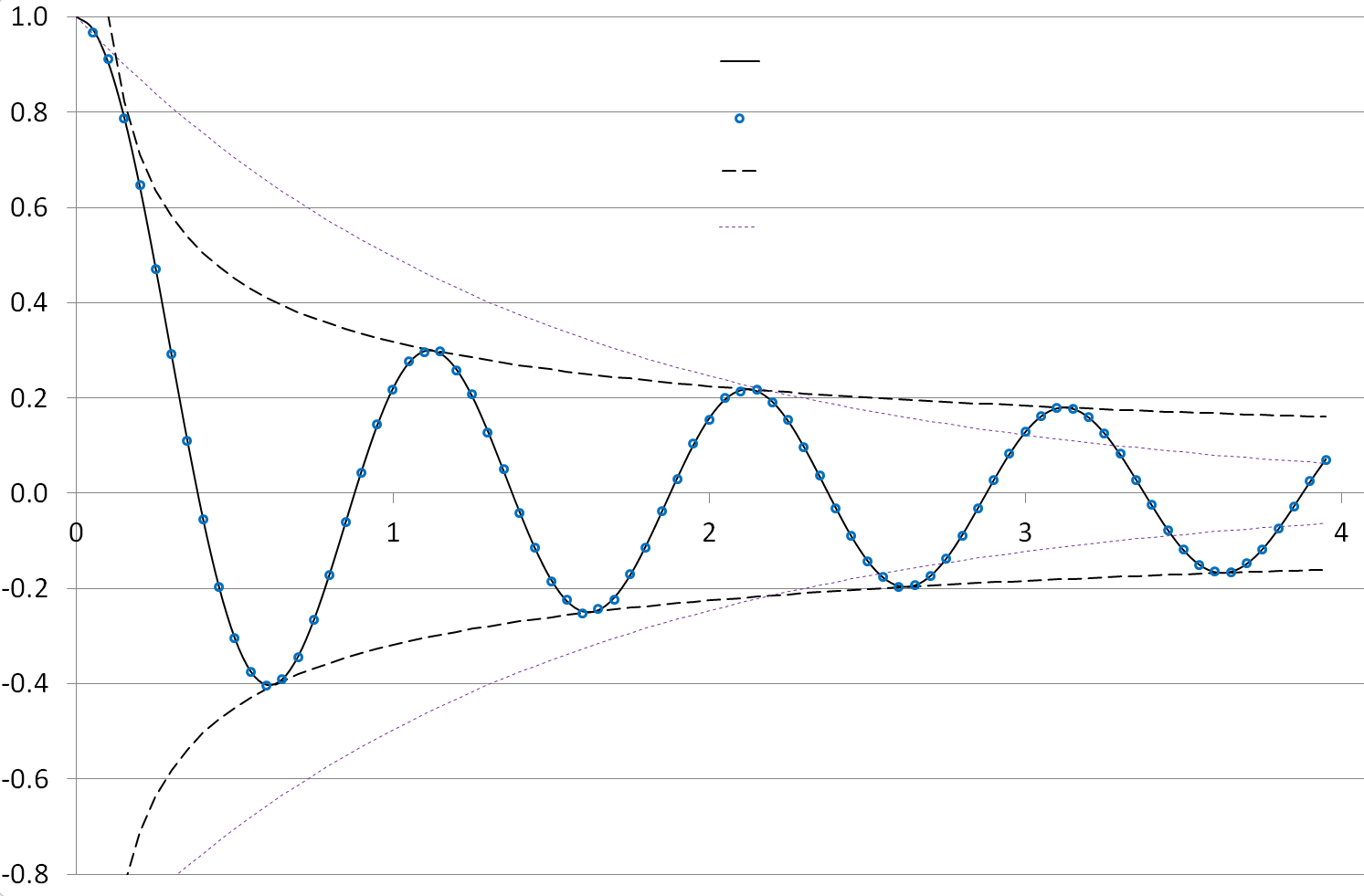}}
\put(150,920){$A(t)/A_0$}
\put(1430,300){$t/t_0$}
\put(845,900){Solution \eq{s1}}
\put(845,839){MD ($10^8$ particles)}
\put(845,782){Envelope $\pm\frac1\pi\sqrt{t_0/t}$}
\put(845,720){Exponential envelope}
\end{picture}
\caption{Oscillational decay of the thermal perturbation amplitude for 1D harmonic crystal. Comparison of the analytical solution \eq{s1} with the MD modeling results ($10^4$ joined chains containing $10^4$ particles each). Dashed lines show the envelope proportional to $1/{\sqrt{t}}$ and also an exponential envelope inherent to the thermal wave equation.}
\end{figure}
The horizontal axis in \fig1 represents the dimensionless time $t/t_0$, where $t_0=\lambda/c$; the vertical axis stands for the oscillation amplitude~$A(t)$, which is computed as the first coefficient of a spatial Fourier expansion of the temperature field. According to \fig1 there is an excellent agreement between the computational results and the analytical curve.

Due to the Bessel function properties~\cite{Abramovits 1979}, the temperature and heat flux~\eq{s1} have an oscillational decay, where the oscillation amplitude is asymptotically proportional to $1/\sqrt t$. The same asymptotics has been obtained in~\cite{Gendelman 2012} for one-dimensional harmonic crystals. In \fig1 the envelope proportional to $1/\sqrt{t}$ is shown by the dashed lines, perfectly bounding both analytical and computational graphs.
The existing theories of heat conduction~\cite{Chandrasekharaiah 1986, Tzou 2015}, such as Fourier's, Maxwell-Cattaneo-Vernotte (MCV), dual-phase-lag~\cite{Tzou 1995}, and spacetime-elasticity~\cite{Gusev-Lurie 2012} yield linear differential equations with constant coefficients, and therefore all of them predict an exponential decay of the sinusoidal perturbation amplitude. In table~1 a comparison of the analytically obtained decay law for $A(t)/A_0$ with the results based on some other theories is demonstrated, an exponential envelope inherent to the thermal wave model is also shown in \fig1. Thus, among the mentioned theories only the current one gives an analytical solution, which agrees with the MD simulations and asymptotic estimations of the oscillation decay for harmonic chains~\cite{Gendelman 2012}.

Let us consider now a stepwise initial temperature distribution, modeling heat transfer between a hot and a cold body:
\be{18}
    x < 0: \quad  T(x) = T_2 \qq
    x > 0: \quad  T(x) = T_1 ,
\ee
where $T_2>T_1$. In this case the integral representations \eq{15}, \eq{17} yield for $|x|\le ct$ an exact analytical solution
\ba{19}{l}
    T(t,x) = T_1 + \frac{\Delta T}{\pi}\arccos{\frac x{ct}}, \\[4mm]
    q(t,x) = \frac{\kB\rho c\Delta T}{\pi}\sqrt{1-\(\frac x{ct}\)^2},
\ea
where $\Delta T = T_2-T_1$; for $x>ct$ the original temperature distribution remains and the heat flux is zero. According to \eq{19} the heat front propagates with the sound speed $c$ and the heat flux through cross-section ${x=0}$ is constant and equal to $\frac1\pi \kB\rho c\Delta T$. In contrast, use of Fourier's law for the same problem gives the heat flux  proportional to $t^{-1/2}$, which is infinite at $t=0$ (an unphysical consequence of Fourier's law). Thus the heat flux $\frac1\pi \kB\rho c\Delta T$ is provided by the temperature difference that is realized on the spatial interval $x\in[-ct,ct]$ with increasing length of $2ct$. Consequently, the heat flux depends on the temperature difference rather than on the temperature gradient. This is in qualitative agreement with the known phenomenon of thermal superconductivity: the heat flux through a one-dimensional harmonic crystal placed between two thermal reservoirs does not depend on the length of the crystal~\cite{Rieder 1967, Dhar 2008}. The same value $\frac1\pi \kB\rho c\Delta T$ was obtained in \cite{Rubin 1971} as a steady-state limit of the heat flux for large $t$.

In \fig2 the analytical solution \eq{19} is compared with computer simulations for $T_2 = 2T_1$.
\begin{figure}[htb]\footnotesize
\unitlength = 1.5\pix
\noindent
\begin{picture}(1360,746)(0,0)
\put(0,0){\includegraphics[width = 86.50mm, height = 47.45mm]{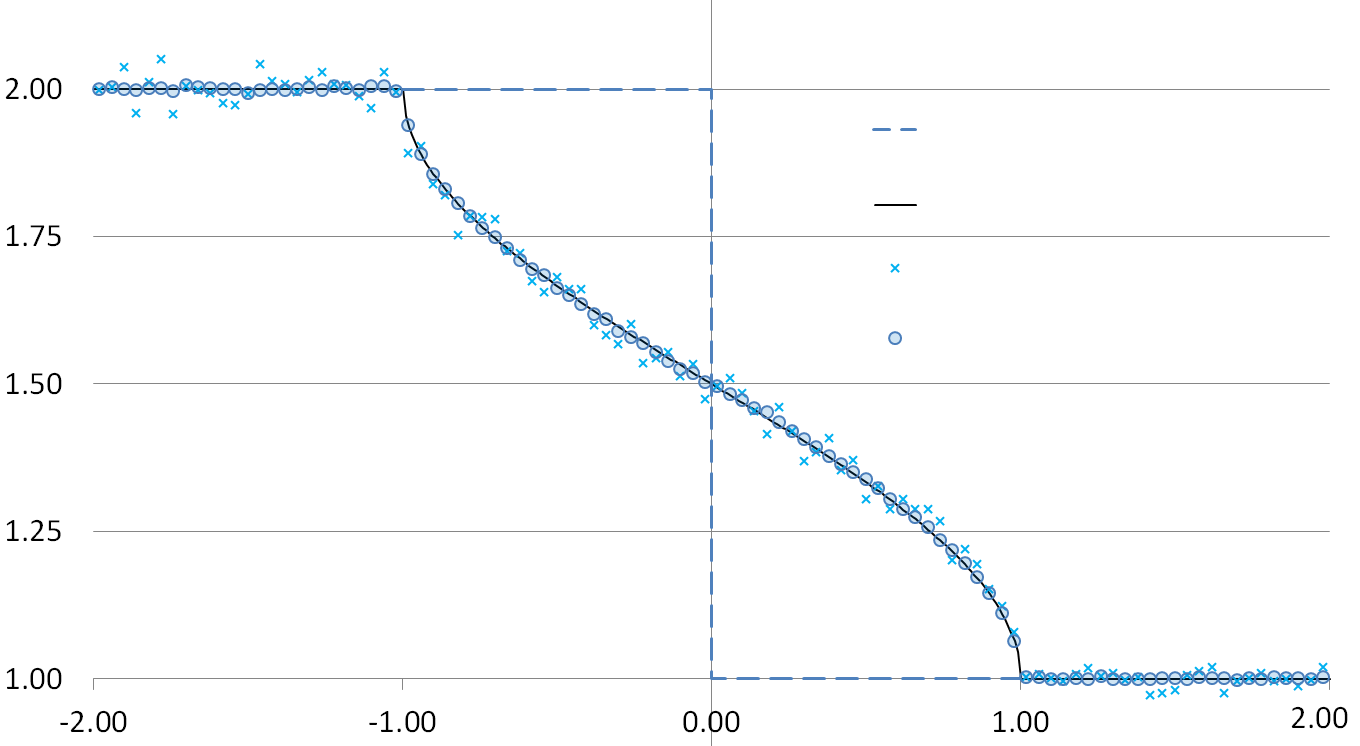}}
\put(0,710){$T/T_1$}
\put(1250,100){$x/ct$}
\put(935,603){Initial temperature}
\put(940,532){Solution \eq{19}}
\put(940,464){MD ($10^6$ particles)}
\put(940,397){MD ($10^8$ particles)}
\end{picture}
\caption{Heat transfer between hot (left) and cold (right) areas of 1D harmonic crystal. The analytical solution \eq{19} is compared with the computer simulation (MD): $10^3$ chains containing $10^3$ particles each (cross is an average over 10 particles); $10^4$ chains containing $10^4$ particles (circle is an average over 100 particles).}
\end{figure}
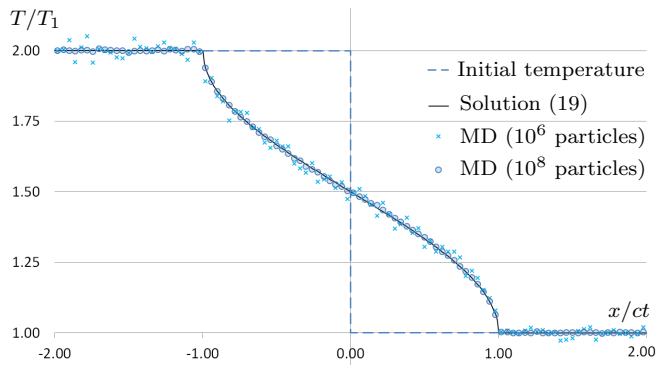
The above described computation procedure is used. \fig2 shows the initial temperature distribution, the analytical solution, and the computation results obtained at~$t=t_0/8$ using $10^6$ and~$10^8$ particles ($t_0=L/c$, where $L$ is the chain length; only half of the chain is shown in the figure). Convergence to the analytical solution with the increase of the system size is clearly seen.

\fig3 shows a part of \fig2 corresponding to positive~$x$. For
symmetry reasons this case can be interpreted as a problem of a half-space heating: the initial temperature for $x>0$ is $T_1$ and the boundary condition at $x=0$ is $T=(T_2+T_1)/2>T_1$.
\begin{figure}[htb]\footnotesize
\unitlength = 1.25\pix
\noindent
\begin{picture}(1591,801)(20,0)
\put(0,0){\includegraphics[width = 84.32mm, height = 42.45mm]{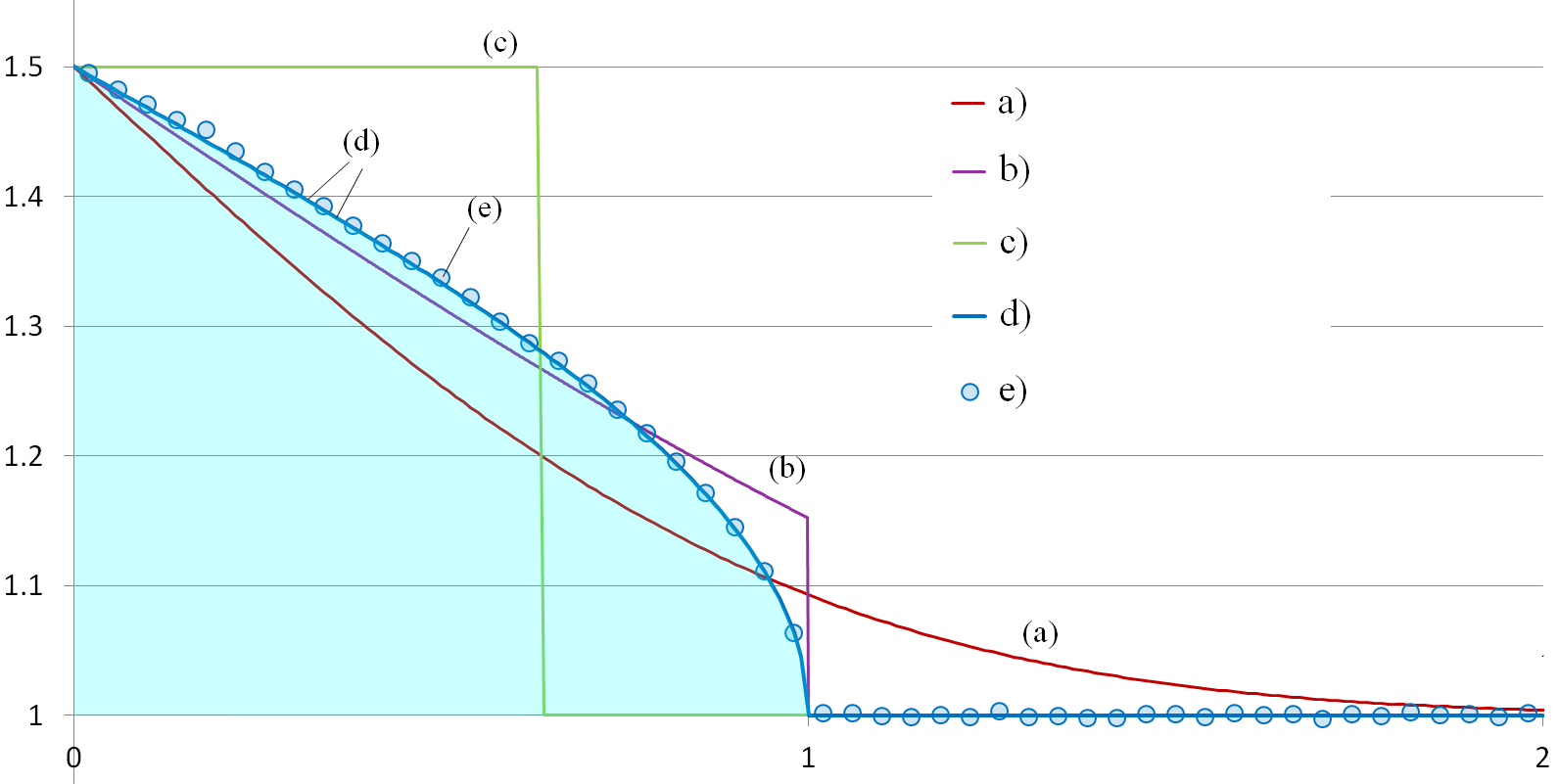}}
\put(10,780){$T/T_1$}
\put(1505,110){$x/ct$}
\put(1060,685){$\dot T = \beta T''$}
\put(1060,618){$\ddot T +\frac1\tau\dot T = \frac\beta\tau T''$}
\put(1060,543){$\ddot T = c^2 T''$}
\put(1060,470){$\ddot T +\frac1t\dot T = c^2 T''$}
\put(1060,390){MD ($10^8$ particles)}
\end{picture}
\caption{Heat propagation for different 1D models: a)~heat equation,
b) thermal wave equation,
c)~wave equation,
d)~equation \eq{13},
e) computer simulation for 1D harmonic crystal ($10^4$ chains containing $10^4$ particles each).}
\end{figure}
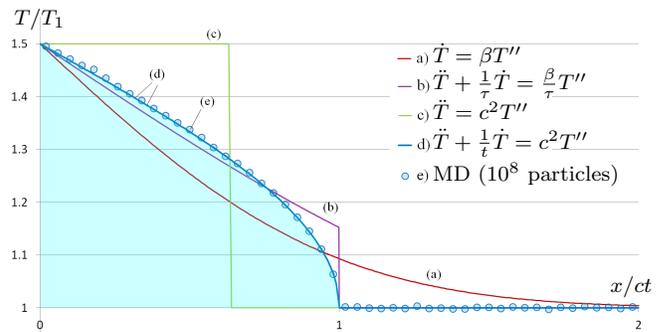
The advantage of this formulation is that the constant boundary temperature is maintained without any thermostat.
This is important since the heat transfer can substantially depend on the thermostat properties~\cite{Dhar 2001, Hoover 2013}. Solutions of the considered problem using four different continuum equations are compared in \fig3 with the simulation results. Parameters are chosen in such a way that the total heat quantity transferred through the cross-section $x=0$ (area under each curve) is equal for all models and the heat front (when it exists) propagates with the sound speed $c$. According to \fig3 the computation results almost coincide with the analytical solution of equation~\eq{13} and significantly differ from the solutions based on the other theories of thermal conduction. Indeed, the classical heat equation predicts no heat front, the thermal wave equation gives a stepwise front, while the real heat front is described by  a smooth curve having a vertical tangent at $x=ct$. Note that the thermal wave (MCV) equation behaves as wave equation at small times and as heat equation at large times~\cite{Babenkov 2014}. However, according to the analytical solution \eq{19} and the presented computer simulations, the heat transfer in a one-dimensional harmonic crystal is self-similar, i.\,e.~$T=T(\frac x{ct})$.
 		
Thus, a nonlocal temperature (a generalization of the kinetic temperature) is introduced in this work to obtain closed description of thermal transfer in a one-dimensional harmonic crystal. Finally this yields to a partial differential equation \eq{13} for the kinetic temperature, which can be referred to as the time-reversible thermal wave equation. The resulting macroscopic constitutive law~\eq{16} (an alternative of Fourier's law for the considered system) predicts a finite velocity of the heat front and independence of the heat flux on the crystal length. The analytical findings are in excellent agreement with the molecular dynamics simulations and previous analytical estimations. The obtained results are relevant to aspects of nanotechnology that involve heat transfer processes in high purity nanostructures~\cite{Chang-Zettl 2008, Xu 2014, Goldstein 2007}.

The author is grateful to
O. V. Gendelman,
W.~G.~Hoover,
D.~A.~Indeitsev,
M.~L.~Ka\-cha\-nov,
V.~A.~Kuzkin,
S.~A.~Lurie,
and N.~F.~Morozov
for helpful and stimulating discussions;
to M.~B.~Babenkov and D.~V.~Tsvetkov for additional analysis and computations.
The work was supported by RSF grant 14-11-00599.

\end{document}